\documentclass[9pt,article,twoside]{rmaa-rho-class/rmaa-rho}

\RMxAAtemplatetype{\RMxAA} 

\def\aap{A\&A\/}
\def\apjs{ApJS}

\newcommand{\angstrom}{\mbox{\ensuremath{\mathring{\mathrm{A}}}}}

\vol{61}
\pages{30-36}
\thisyear{2026}
\doi{\href{https://www.astroscu.unam.mx/rmaa/RMxAA..XX-X}{https://www.astroscu.unam.mx/rmaa/RMxAA..XX-X}}

\title{Mapping the Quasar Main Sequence in the UV range: A Connection with the UV Fe III Emission}

\author[1,2,3]{D.~ Martínez Collipal \orcidlink{0009-0004-4993-5656}}
\author[1,2,3]{M.~L.~Martínez-Aldama \orcidlink{0000-0002-7843-7689}}
\author[4]{A.~Pandey \orcidlink{0000-0003-3820-0887}}
\author[1]{M.~Cardenas}
\author[5,6]{P.~Marziani \orcidlink{0000-0001-6208-9109}}
\author[,7]{S.~Panda \orcidlink{0000-0002-5854-7426}\thanks{Gemini Science Fellow}}

\affil[1]{Astronomy Department, Universidad de Concepción, Casilla 160-C, Concepción 4030000, Chile}
\affil[2]{ Millennium Nucleus on Transversal Research and Technology to Explore Supermassive Black Holes (TITANs)}
\affil[3]{ Millennium Institute of Astrophysics (MAS), Nuncio Monseñor Sótero Sanz 100, Providencia, Santiago, Chile }
\affil[4]{Department of Physics and Astronomy, University of Utah, Salt Lake City, UT 84112, USA}
\affil[5]{National Institute for Astrophysics (INAF), Astronomical Observatory of Padua, Vicolo dell'Osservatorio 5, IT 35122, Padova, Italy \quad}
\affil[6]{Instituto de Astrof\'\i sica de Andaluc\'\i a (IAA--CSIC), Glorieta de Astronomia s/n, ES 18008 Granada, Spain \quad}
\affil[7]{International Gemini Observatory/NSF NOIRLab, Casilla 603, La Serena, Chile}

\leadauthor{D. Martínez Collipal et al.}
\smalltitle{LaTex Macro RMxAA}

\corres{Diego Martínez Collipal}
\email{diemartinez2021@udec.cl}

\received{January 30, 2026}
\accepted{\today}

\license{Texto de la licencia aquí}

\setbool{rho-abstract}{true} 
\setbool{rho-resumen}{true} 

\begin{abstract}
The quasar main sequence (MS) provides a consistent unification scheme for type I AGNs. This framework, constructed from the relative flux between the optical emissions $\mathrm{H}\beta$ and $\ion{Fe}{ii}\,(\mathrm{\lambda}4570)$, and the $\mathrm{FWHM(H\beta)}$, organizes quasars into two populations (B, A) and the subpopulation xA, which typically consists of highly accreting sources. We explored the use of the UV 1900 blend emissions ($\ion{Al}{iii}\,\mathrm{\lambda}1860$, $\ion{Si}{iii}]\,\mathrm{\lambda}1892$, $\ion{C}{iii}]\,\mathrm{\lambda}1909$) and the $\ion{Fe}{iii}$ and $\ion{Fe}{ii}$ transitions between 1700 and 2200$\,\angstrom$ as tracers of this main sequence. We fitted these emissions for a sample of 69 low-$z$ ($z\lesssim0.8$) quasars observed with the Hubble Space Telescope (HST) that had available optical measurements, enabling a direct comparison with the optical quasar main sequence. We found a UV plane defined by $\mathrm{FWHM}(\ion{Al}{iii})$ and the equivalent width of the combined emissions of $\ion{Fe}{iii}$ and $\ion{Fe}{ii}$. The plane defined by these parameters mirrors the quasar main sequence in the optical regime and shows a clear distinction between the different populations. The plane also enables the identification of xA sources with high completeness but a non-negligible rate of false positives. Our results suggest that the UV iron emission around the 1900 blend could be used to trace the quasar main sequence trend, offering an alternative way to study this scheme at high redshift. 
\end{abstract}


\begin{resumen}
La secuencia principal de cuásares constituye un esquema unificador para los AGNs de tipo I. Este esquema, construido a partir de la razón de flujos entre las emisiones ópticas de $\mathrm{H}\beta$, $\ion{Fe}{ii}\,(\mathrm{\lambda}4570)$ y el $\mathrm{FWHM(H\beta)}$, organiza a los cuásares en dos poblaciones (B, A) y la subpoblación xA, la cual típicamente consiste de objetos con alta acreción. Investigamos el uso de las emisiones ultravioletas del blend en 1900; $\ion{Al}{iii}\,\mathrm{\lambda}1860$, $\ion{Si}{iii}]\,\mathrm{\lambda}1892$, $\ion{C}{iii}]\,\mathrm{\lambda}1909$ y las transiciones de $\ion{Fe}{iii}$ y $\ion{Fe}{ii}$ entre 1700 y 2200$\,\angstrom$, como trazadores de esta secuencia principal. Ajustamos estas emisiones en 69 cuásares de bajo $z$ ($z \lesssim 0.8$) observados con el Telescopio Espacial Hubble (HST), los cuales tienen mediciones en el óptico disponibles, permitiendo una comparación directa con la secuencia principal en el óptico. Encontramos un plano UV definido por $\mathrm{FWHM}(\ion{Al}{iii})$ y el ancho equivalente de la emisión combinada de $\ion{Fe}{iii}$ y $\ion{Fe}{ii}$. El plano UV definido por estos parámetros se asemeja a la secuencia principal de cuásares en el óptico y muestra una clara distinción entre las diferentes poblaciones. Este plano también permite la identificación de los objetos xA, con una alta completitud, pero un ratio no despreciable de falsos positivos. Nuestro resultado sugiere que la emisión UV de hierro alrededor del 1900 blend podría ser usada para trazar la secuencia principal de cuásares, ofreciendo una alternativa para estudiar este esquema a alto corrimiento al rojo. 
\end{resumen}

\begin{document}

\maketitle
\pagestyle{fancy}\thispagestyle{firststyle}


\section{INTRODUCTION}

    \RMxAAstart{T}
    he Eigenvector 1 or Quasar Main Sequence (MS) of \citet{Boroson1992} is a powerful unification scheme for Type I AGNs \citep{Sulentic2000,Zamfir2010,Kuraszkiewicz2009,Marziani2018,Panda2024FrASS..1179874S}. This diagram is constructed with the $\mathrm{FWHM}(\mathrm{H}\beta)$, which represents the width measured at 50\% of the $H\beta$ line intensity, and $\mathrm{R}_{\rm FeII}$, which is the flux ratio between $\mathrm{H}\beta$ and the $\ion{Fe}{ii}$ emission centered at $4570\,\angstrom$.
    The MS groups type I AGNs into two populations: Population B (Pop. B), with FWHM(H$\beta$) $\gtrsim$ 4000\,km\, s$^{-1}$, and Population A (Pop. A), with FWHM(H$\beta$) $\lesssim$ 4000\,km\, s$^{-1}$ \citep{Sulentic2000}. A relevant subset of Pop. A is the "Extreme population A" or "xA", which are the objects that satisfy the criterion: $\mathrm{R}_{\mathrm{\rm FeII}} > 1.0$ \citep{Marziani2014,Du2016,Negrete2018,Panda2019ApJ...882...79P}. The existence of different populations is currently explained by the increase in the Eddington ratio from Pop. B to xA, which modifies the physical conditions in the broad line region (BLR), thereby changing the observed spectra \citep{Marziani2025}. In addition, the black hole mass and viewing angle also influence the MS trend \citep{Marziani2003,Shen2014,Panda2019ApJ...882...79P,Naddaf2025}.\\
The MS classification scheme has been widely explored in the local Universe and up to cosmic noon \citep{Sulentic2000,Marziani2003,Marziani2009, Martínez-Aldama2021,Deconto-Machado2023, Buendia-Rios2025}. At higher redshifts, observations of H$\beta$ and Fe\,II are affected by telluric absorption and sky emission \citep{Marziani2025}. Consequently, few sources have been explored at $z > 3$ \citep{Yang2023,Loiacono2024,Trefoloni2025}, and a proper study of the MS at these redshifts is still needed. In this context, the use of ultraviolet (UV) analogues of H$\beta$ and Fe\,II can be an alternative, since this range is observable with ground-based facilities and large surveys such as SDSS until $z \sim 4$.\\ 
    We searched for these analogues among the UV emissions within $1700-2200\,\angstrom$. We focused our attention on two features, the $\lambda1900$ blend and the underlying $\ion{Fe}{iii}$ and $\ion{Fe}{ii}$ pseudo-continuum. The $\lambda1900$ blend is the combined emission of three emission lines: $\ion{Al}{iii}\,\mathrm{\lambda}1860$, $\ion{Si}{iii}]\,\mathrm{\lambda}1892$, and $\ion{C}{iii}]\,\mathrm{\lambda}1909$. The relative intensity between these lines changes along the MS. In Pop. B, $\ion{C}{iii}]\,\mathrm{\lambda}1909$ dominates the feature, while in Pop. A $\ion{Al}{iii}\,\mathrm{\lambda}1860$, $\ion{Si}{iii}]\,\mathrm{\lambda}1892$ are enhanced \citep{Bachev2004, Negrete2012, Marziani2014}. The underlying UV iron pseudo-continuum is produced by the multiple transitions of $\ion{Fe}{iii}$ and $\ion{Fe}{ii}$ emitted in this range. The current evidence about this feature suggests that the $\ion{Fe}{iii}$ emission is particularly strong in xA quasars \citep{Vestergaard2001,Martínez-Aldama2018,Temple2020}. \\
Motivated by these findings, we explored whether the UV $\ion{Fe}{iii}$ emission can be a proxy for the optical $\ion{Fe}{ii}$ and whether the lines of the $\lambda1900$ blend can replace H$\beta$ to reproduce the quasar main sequence using UV tracers. \\
This work is organized as follows. In Section \ref{sec:sample}, we define the properties of our studied sample. Section \ref{sec:methods} describes our technique to fit the studied UV emissions. Section \ref{sec:results} presents the main results. Finally, we discuss the main conclusions and caveats in Section \ref{sec:conclusion}.
\section{Sample}
\label{sec:sample}
Our analysis is based on a subsample of 69 HST AGNs at $z<0.8$ from \citet{Sulentic2007}. These sources were selected for their coverage of the $1700-2200\angstrom$ range, allowing for the simultaneous analysis of the $\lambda1900$ blend and the UV iron emission. To classify the sample within the MS trend, we adopted the values of $\mathrm{FWHM(H\beta})$ and $\mathrm{R_{\rm FeII}}$ presented in \citet{Sulentic2007}, and for a few sources, the updated values of \citet{Floris2024}. The details can be found in an ASCII table\footnote{\url{https://github.com/DiegoMartinez-astro/HST-FOS-Fe-III-emission/tree/main}}. The studied sample is balanced in terms of the MS populations, comprising 56\% for Pop. B and 44\% for Pop. A. The xA sources account for 7\% of the sample, which is close to the typical proportion of this class in large samples \citep[10\%,][]{Zamfir2010,Shen2011}. The bolometric luminosity in our sample ranges from $\mathrm{L}_{\mathrm{bol}} \sim 10^{44}\,\mathrm{erg\,s^{-1}}$ to $\mathrm{L}_{\mathrm{bol}} \sim 10^{47}\,\mathrm{erg\,s^{-1}}$, with a median value of $\mathrm{L}_{\mathrm{bol}} \sim 10^{46}\,\mathrm{erg\,s^{-1}}$. We estimated these values using the continuum flux at $1700\,\angstrom$, following \citet{Marziani2022} prescription. \\
We characterized the signal-to-noise ratio (SNR) of each spectrum using a continuum window of $10\,\angstrom$ in width centered at $1700\,\angstrom$. We found a median SNR of $ \sim 24$ and a small fraction (10\%) of observations with low quality (SNR$ < 10$).
\section{Methods}
\label{sec:methods}
We used a multicomponent fitting code to model the $1700-2200\,\angstrom$ range, which includes the emission lines $\ion{N}{iii}]\,\lambda1750$, $\ion{Si}{ii}\,\lambda1816$, $\ion{Al}{iii}\,\lambda1860$, $\ion{Si}{iii}]\,\lambda1892$, and $\ion{C}{iii}]\,\lambda1909$, along with the $\ion{Fe}{iii}$ and $\ion{Fe}{ii}$ pseudo-continuum. The code utilizes the Markov Chain Monte Carlo method (MCMC) to explore the parameter space of all components. Subsequently, the best-fit model is obtained through proper $\chi^{2}$ minimization \citep{Barlow1989}. To ensure an accurate decomposition, we first determined the underlying continuum manually over the extended range $1400-2200\,\angstrom$ using a power-law model \citep{Malkan1982}. This continuum was then fixed during the multicomponent fitting code.\\
We modeled all lines with a single broad component (BC), and we additionally included a narrow component for $\ion{C}{iii}]$. The $\ion{Al}{iii}$ doublet was modeled with two BCs centered at $\lambda1854$ and $\lambda1862$, with a fixed intensity ratio of 1.25:1, following \citet{Laor1997}. Regarding the line profiles, we followed the prescription of \citet{Negrete2012}, and we modeled $\ion{Al}{iii}$ and $\ion{Si}{iii}]$ with Gaussian profiles in Pop. B sources, and Lorentzians in Pop. A. The remaining lines were modeled exclusively with Gaussian profiles. In the code, the FWHM of $\ion{Si}{iii}]$ was fixed to match that of $\ion{Al}{iii}$ \citep{Negrete2012}. Conversely, we relaxed the constraints on the remaining FWHMs, fluxes, and shifts to let them converge naturally. \\
The UV iron pseudo-continuum was modeled with the templates of \citet{Bruhweiler2008} for $\ion{Fe}{ii}$ and \citet{Vestergaard2001} for $\ion{Fe}{iii}$. These templates were rescaled and broadened to match the observed emission. The broadening value was selected to match the $\mathrm{FWHM}$ of $\ion{Al}{iii}$, or determined manually when this choice did not yield a satisfactory fit. Finally, we included additional Gaussian components at $1715\,\angstrom$, $1785\,\angstrom$, $1914\,\angstrom$, $2000\,\angstrom$, $2020\,\angstrom$, and $2080\,\angstrom$, following \citet{Martínez-Aldama2018} prescription. 
\section{Results}
\label{sec:results}
\subsection{Exploring line ratios based on FeIII}
\begin{figure}
\centering
\captionsetup{width=\linewidth}
\captionsetup{font=small}
\includegraphics[width=0.95\columnwidth]{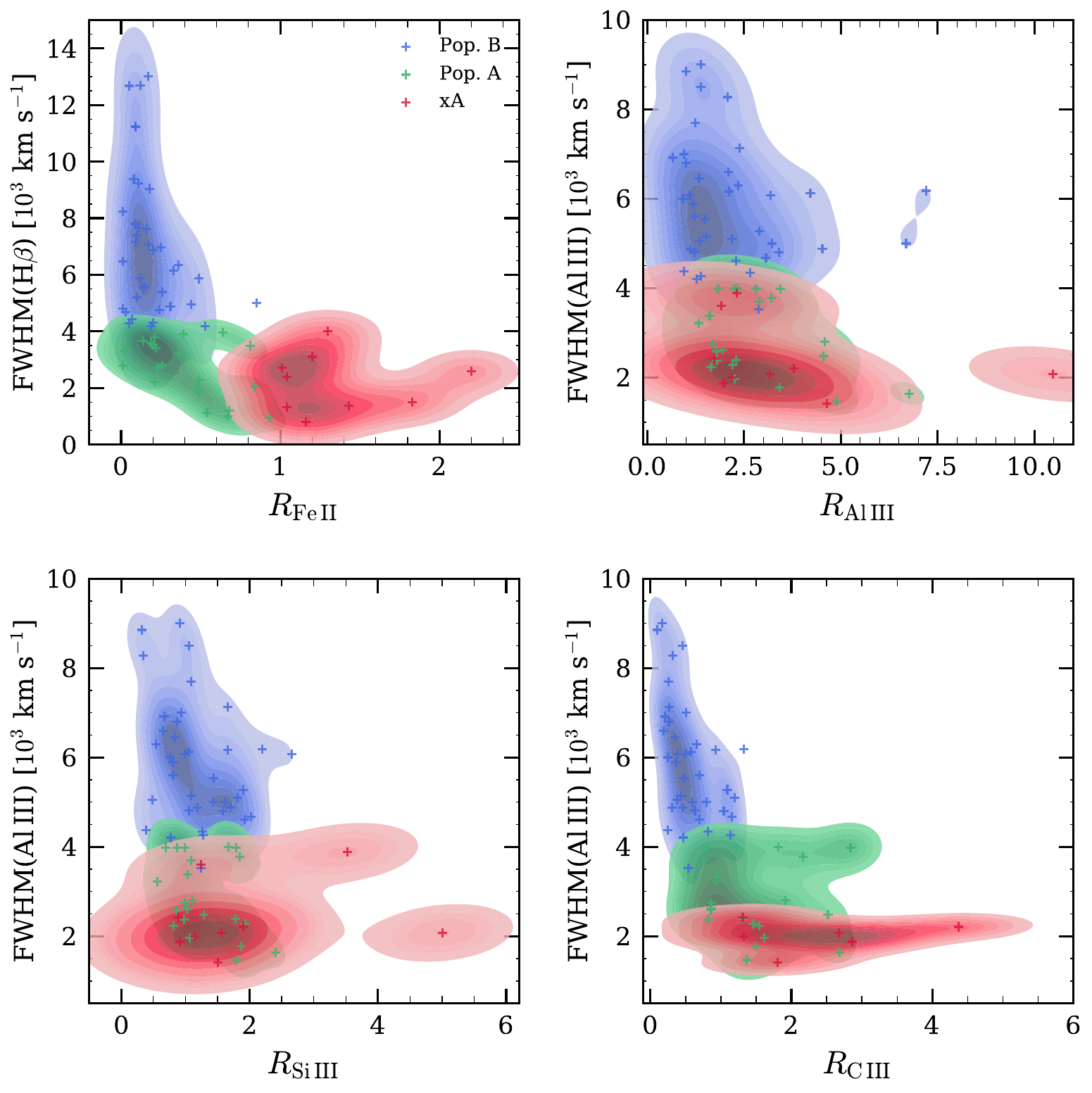}
\caption{\justifying Top left panel shows the Optical Main sequence of our sample, with each object color-coded according to its population. We additionally included contour density plots, which show 10 contour levels, limited to the $2\sigma$ confidence interval for each population.  The remaining panels keep the same format and present the tested UV diagrams using $FWHM(\ion{Al}{iii})$ and the following flux ratios: Top right: $Flux(\ion{Fe}{iii}+\ion{Fe}{ii})/{Flux(\ion{Al}{iii})}$, bottom left: $Flux(\ion{Fe}{iii}+\ion{Fe}{ii})/{Flux(\ion{Si}{iii}])}$, bottom right: $Flux(\ion{Fe}{iii}+\ion{Fe}{ii})/{Flux(\ion{C}{iii}])}$. In this last panel, three xA sources fall outside the plotted range as a consequence of the near absence of $\ion{C}{iii}]$ in their spectra.}
\label{fig:fe3_ratios}
\end{figure}

To establish a UV analogue of the quasar main sequence, we searched for substitutes for $R_{\rm FeII}$ and FWHM(H$\beta$). For FWHM(H$\beta$), we adopted the FWHM($\ion{Al}{iii}$) as a proxy, motivated by the strong correlation between both parameters \citep{Marziani2022,Dultzin2020}. To replace $R_{\rm FeII}$, we tested different flux ratios between the UV iron emission and the lines of the $\lambda1900$ blend. The flux of this UV iron emission corresponds to the integrated flux in the $1980-2120\,\angstrom$ range, which is isolated from non-iron features. The integration includes $\ion{Fe}{iii}$ and $\ion{Fe}{ii}$ transitions, since we were unable to accurately separate them. However, the contribution of $\ion{Fe}{ii}$ is expected to be negligible in this range \citep{Vestergaard2001}. The tested line ratios are defined as follows:   
\begin{itemize}
    \item $\mathrm{R_{Al\,III}=Flux(\ion{Fe}{iii}+\ion{Fe}{ii})/{Flux(\ion{Al}{iii})}}$
    \item $\mathrm{R_{Si\,III}=Flux(\ion{Fe}{iii}+\ion{Fe}{ii})/{Flux(\ion{Si}{iii})}}$
    \item $\mathrm{R_{C\,III}=Flux(\ion{Fe}{iii}+\ion{Fe}{ii})/{Flux(\ion{C}{iii})}}$ 
\end{itemize}
Using these ratios and the FWHM($\ion{Al}{iii}$), we constructed MS-like diagrams. In these diagrams, we color-coded the objects according to their optical MS classification, and we drew contour density plots to visualize the organization of the MS. In each diagram, we analyzed whether the studied ratio is able to reproduce the MS organization or if, on the contrary, it is unable to separate the different classes. The results are presented in Figure \ref{fig:fe3_ratios}.\\
The resulting diagrams indicate that $\mathrm{R_{Al\,III}}$ and $\mathrm{R_{Si\,III}}$ are insufficient to distinguish between Pop. A and xA. This is probably a consequence of the strong $\ion{Al}{iii}$ and $\ion{Si}{iii}]$ emission in the xA regime \citep{Bachev2004}. As a result, when the UV iron emission is normalized by these lines, the xA sources shift toward lower values on the x-axis, and the contour plots tend to overlap.  $\mathrm{R_{C\,III}}$ seems to be a better option than previous ratios. However, we emphasize that this ratio reaches out-of-scale values in objects with the highest $R_{\rm FeII}$, since they show almost absent $\ion{C}{iii}]$ emission \citep{Negrete2012}. These objects were not considered in the contour plot, since they significantly distorted the figure and fell outside of the plotted range. 
\subsection{Equivalent width approach}
\begin{figure}
\centering
\captionsetup{width=\linewidth}
\captionsetup{font=small}
\includegraphics[width=0.95\columnwidth]{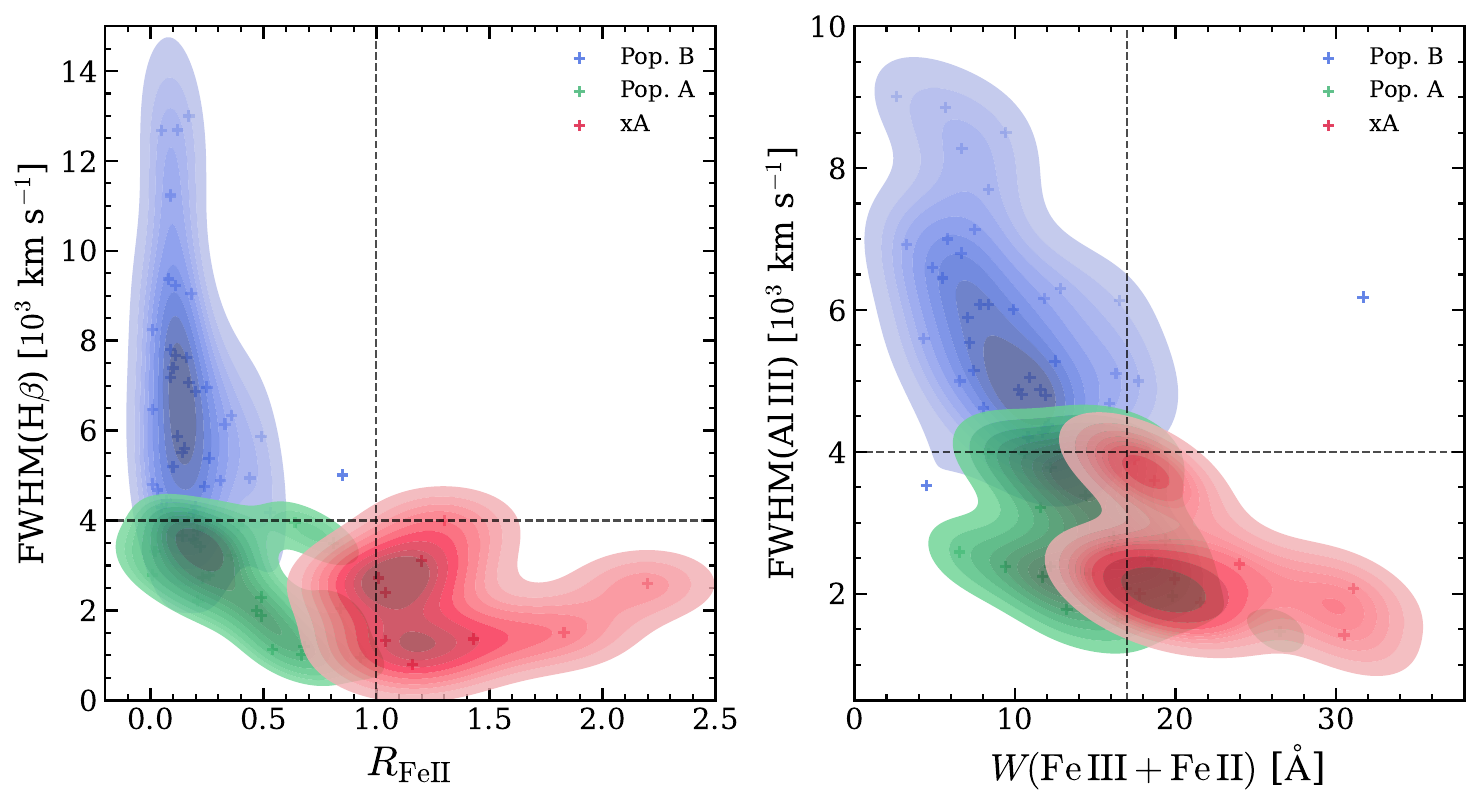}
\caption{\justifying
Left: Optical main sequence of our sample. The grey lines mark the limits between populations B, A, and xA. Right: $FWHM(\ion{Al}{iii})$ against the equivalent width of the UV iron emission. In both panels, each object is color-coded according to its optical MS classification. We included contour density plots, which show 10 contour levels, limited to the $2\sigma$ confidence interval for each population. The grey lines on the right panel show the tentative limits that we derived between the MS populations.}
\label{fig:uv_ms}
\end{figure}

In this section, we explored a different approach using the equivalent width (W) of the UV iron emission as a proxy for $R_{\rm FeII}$. These W values were calculated by dividing the UV iron fluxes by the underlying continuum fluxes across the $1980-2120\,\angstrom$ range. The resulting diagram using W is presented in Figure \ref{fig:uv_ms}.
The diagram with equivalent width shows a clear similarity with the optical quasar MS. In this UV plane, each MS population occupies a distinct region. Although some overlap is observed between Pop. A and xA, this is largely an effect of the $2\sigma$ contours rather than an intrinsic overlap. We defined the separation between the MS populations using the Following tentative limits: 
$\mathrm{FWHM(\ion{Al}{iii})}\approx4000$\,km\,s$^{-1}$, between Pop. B and Pop. A. The boundary between Pop.A and xA is instead placed at $\mathrm{W(\ion{Fe}{iii}+\ion{Fe}{ii}}\approx 17\,\angstrom$. The $\mathrm{FWHM(\ion{Al}{iii})}=4000$\,km s$^{-1}$ threshold matches the value that separates Pop. B from Pop. A using H$\beta$ in the optical MS \citep{Sulentic2002}. This division is interpreted as a change in the accretion mode, from a thin accretion disk \citep{Shakura1973} in Pop. B to a slim-disk geometry \citep{Abramowicz1988} in Pop. A \citep[see Figure 1 in ][]{Panda2023Univ....9..492P}. 
The match between the limits in $\ion{Al}{iii}$ and $\mathrm{H\beta}$ supports the strong correlation between the FWHM of these lines, which have been claimed to be equivalent tracers of the virial motion in the BLR \citep{Marziani2022,Dultzin2020,Martínez-Aldama2018}. On the other hand, we found that the threshold $\mathrm{W(\ion{Fe}{iii}+\ion{Fe}{ii})}\sim17\,\angstrom$ is the value that maximizes the correct classification of xA sources (100\%) while maintaining a minimal false-positive rate  (22.7\%). Consequently, it is the optimal cutoff point to distinguish between Pop. A and xA in our diagram. This threshold has no precedent in the literature, and further analysis is needed to clarify its physical origin. \\
The good agreement found between the diagrams in Figure \ref{fig:uv_ms} suggests that $\mathrm{W(\ion{Fe}{iii}+\ion{Fe}{ii})}$ is a better proxy for $R_{\rm FeII}$ than any of the studied line ratios.
\subsection{The xA region in the UV plane}
\begin{figure}
\centering
\includegraphics[width=0.9\columnwidth]{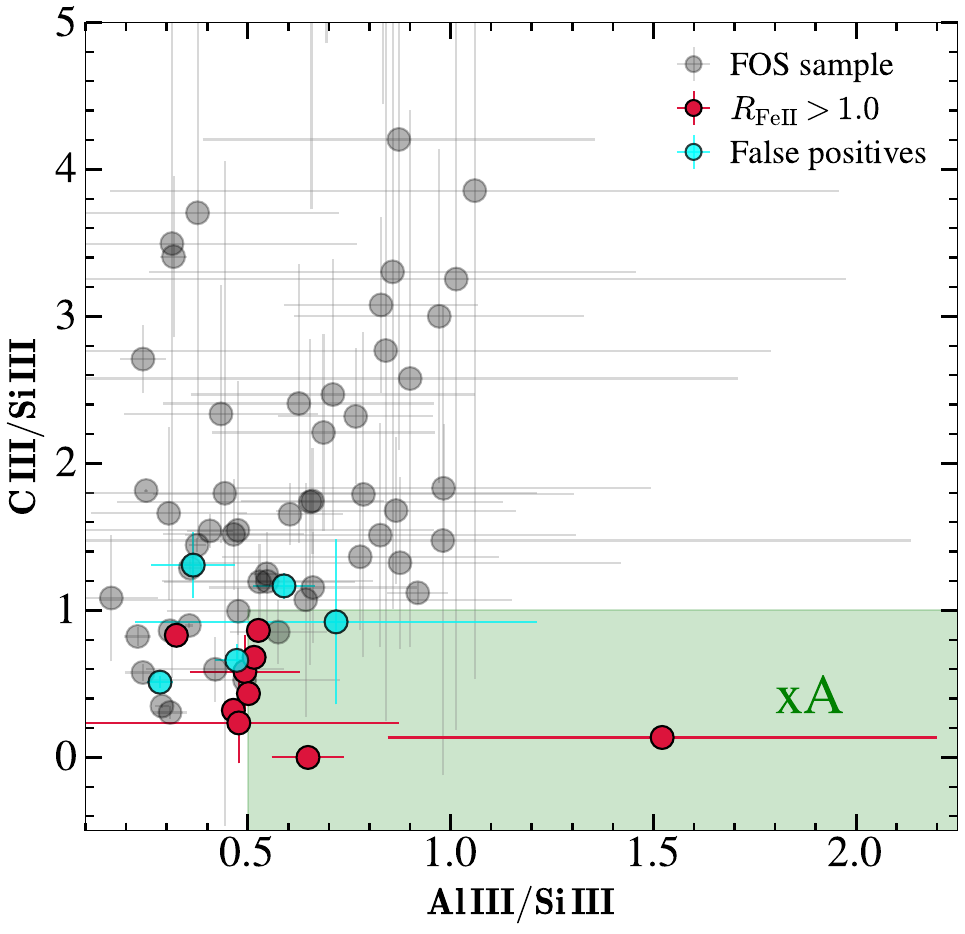}
\caption{\justifying Identification of xA sources in the UV range. The shaded green region corresponds to the criteria of \citet{Marziani2014}. The red circles indicate all objects that satisfy the $R_{FeII} > 1.0$ criterion (i.e., xA sources), and the grey circles are the rest of the HST sample. We highlight with cyan color the position of the false positives produced by the UV iron criteria, as shown in Table \ref{tab:xA}.}
\label{fig:ratios_blend}
\end{figure}

The condition $W(\ion{Fe}{iii}+\ion{Fe}{ii}) \gtrsim 17\,\angstrom$ can be seen as a new criterion to identify xA sources. In the past, \citet{Marziani2014} (M\&S+2014) already developed the criteria $\ion{Al}{iii}\lambda1860/\ion{Si}{iii}]\lambda1892 > 0.5$, $\ion{C}{iii}]\lambda1909/\ion{Si}{iii}]\lambda1892 < 1.0$, which enable the correct identification of xA sources in 80\% of cases \citep{Marziani2014_b,Marziani2022,Buendia-Rios2023}. However, the use of these criteria requires an accurate decomposition of the $\lambda1900$ blend, which is not always achievable in low-signal-to-noise spectra. Conversely, our criterion relies on the fitting of an isolated region, which can be easily achieved using multiple Gaussians without requiring a detailed decomposition.\\
We compared the criteria of M\&S+2014 against our UV iron criterion. To this end, we first needed to identify the true xA sources to have a reference for comparison.  We did this by considering the division provided in the left panel of  Figure \ref{fig:uv_ms}, which is based on the optical measurements of \citet{Sulentic2007} and \citet{Floris2024}. After that, we used our measurements of the 1900 blend to calculate the values of $\ion{Al}{iii}\lambda1860/\ion{Si}{iii}]\lambda1892$ and $\ion{C}{iii}]\lambda1909/\ion{Si}{iii}]\lambda1892$. We applied the M\&S+2014 criteria to these values to see how many true xA sources were recovered. This result is presented in  Figure \ref{fig:ratios_blend}. The figure shows that the xA sources in the diagram lie close to the boundary of the defined xA region, which means that they barely satisfy the criteria. In addition, some false positives are also observed.\\
We compared the rate of correctly identified xA sources by the M\&S+2014 criteria with our UV iron criterion $W(\ion{Fe}{iii}+\ion{Fe}{ii}) \gtrsim 17\,\angstrom$, which is summarized in Table \ref{tab:xA}.
\begin{table}[t]
\centering
\caption{Comparison of different xA selection criteria.}
\label{tab:xA}
\renewcommand{\arraystretch}{1.3}
\setlength{\tabcolsep}{0pt} 
\begin{tabular*}{\columnwidth}{@{\extracolsep{\fill}}lcccc}
\toprule
Method & TP & FP & Completeness & Purity \\
\midrule
UV \ion{Fe}{iii} & 9 & 5 & 100\% & 64.3\% \\
M\&S+2014 & 5 & 2 & 55.6\% & 71.4\% \\
\bottomrule
\end{tabular*}
\begin{flushleft}
\footnotesize 
\textbf{Notes:} \\
\textit{Methods:} \textbf{UV Fe\,III} uses $W(\ion{Fe}{iii}+\ion{Fe}{ii}) \geq 17\,\angstrom$. \textbf{M\&S+2014} uses $\ion{Al}{iii}\lambda1860/\ion{Si}{iii}]\lambda1892 > 0.5$, $\ion{C}{iii}\lambda1909/\ion{Si}{iii}]\lambda1892 < 1.0$.
\textit{Parameters:} \textbf{TP}: Correctly identified xA; \textbf{FP}: non-xA included; \textbf{Completeness}: Fraction of total xA sources recovered; \textbf{Purity}: Fraction of selected objects that are true xA.
\end{flushleft}
\end{table}
The metrics show that the M\&S+2014 criteria provide a more conservative approach, with a reduced number of false positives, but lower completeness. In contrast, our UV iron criterion recovers all xA sources, but with a higher false-positive rate of 22.7\%. To better understand the nature of these false positives, we analyzed their distribution in Figure \ref{fig:ratios_blend}. One of the false positives is also misclassified as xA by the M\&S+2014 criteria, and two others appear close to the boundary. Suggesting that their nature might be similar to those of an optically selected xA. Only two false positives are clearly inconsistent with the M\&S+2014 criteria. \\
The existence of multiple borderline sources and false positives in the diagram likely reflects that xA sources are a continuous extension of Pop. A, rather than a distinct class. This becomes more evident in our xA sample, as their UV spectral features do not differ significantly from those of Pop. A. Our xA objects typically show strong $\ion{Si}{iii}]\lambda1892$ but a modest amount of $\ion{Al}{iii}\lambda1860$, making them less extreme than other xA samples of the literature (see \citealt{Martínez-Aldama2018}). In this context, the high completeness but modest purity of our UV iron criterion suggests that it might be a necessary but not sufficient condition to identify an xA source. Nevertheless, both methods have comparable metrics, and the straightforward application of the UV iron criterion makes it an interesting alternative for further study. 
\section{Conclusions and Discussions}
\label{sec:conclusion}
We performed a spectral decomposition of 69 low-z AGN observed with the HST at $1700-2200\,\angstrom$, exploring their evolution along the quasar main sequence, with special interest in the $\lambda1900$ blend and $\ion{Fe}{iii}$ emission. We tested different line ratios as proxies for $R_{\rm FeII}$ using $\ion{Fe}{iii}$ and the blend lines. We concluded that the equivalent width of the combined UV $\ion{Fe}{iii}$+$\ion{Fe}{ii}$ emission is the best proxy for $R_{\rm FeII}$. We defined a UV plane using $\mathrm{FWHM(\ion{Al}{iii})}$ and $\mathrm{W(\ion{Fe}{iii}+\ion{Fe}{ii})}$ that mirrors the quasar MS, effectively reproducing the trend along the optical MS and the distribution of its populations (see Figure \ref{fig:uv_ms}). We used this UV plane to propose a straightforward methodology for identifying xA sources, which employs $\mathrm{W(\ion{Fe}{iii}+\ion{Fe}{ii})} \sim  17\, \angstrom$ as a lower limit to isolate these objects. We compared the precision of this technique against the criteria of \citet{Marziani2014}, and found comparable results. Nevertheless, we emphasize that our criteria produce a significant rate of false positives (22.7\%).\\ In summary, our results suggest the following UV criteria to organize quasars. Pop. B and Pop. A can be separated using $\mathrm{FWHM(\ion{Al}{iii})}\approx 4000\, km\, s^{-1}$ as a boundary, analogous to the boundary defined by H$\beta$  in the optical plane. xA quasars can be broadly separated from Pop. A through their stronger $\ion{Fe}{iii}+\ion{Fe}{ii}$ emission. We found that $\mathrm{W(\ion{Fe}{iii}+\ion{Fe}{ii})} \sim  17\, \angstrom$ is a suitable threshold to separate both populations. \\
We conclude that the UV $\ion{Fe}{iii}$ and $\ion{Fe}{ii}$ emission contains valuable information that should be considered when studying the $\lambda$1900 blend. We found that this UV iron emission varies along the quasar main sequence, which suggests similarities in the excitation mechanism of $\ion{Fe}{iii}$ in the UV and $\ion{Fe}{ii}$ in the optical. Further modeling with photoionization simulations would be helpful to corroborate this idea.\\
We emphasize that our sample is limited and might not capture the full variety of sources observed in the MS. We are especially concerned about the application of our methodology to higher luminosity samples, since our studied xA sources are not as luminous as other objects in the literature.\\
Despite these limitations, our work presents a new manifestation of the MS trend in the ultraviolet range, which has the advantage of being observable at high redshifts (z > 5), potentially extending our understanding of the quasar main sequence into the early Universe. 

\renewcommand{\refname}{REFERENCES}
\bibliography{rmaa}



\section{ACKNOWLEDGEMENTS}
DMC acknowledges the funding from the National Agency for Research and Development (ANID), Chile, through the postgraduate scholarship ANID BECAS/MAGISTER NACIONAL, 22251176. MLMA and DMC acknowledge financial support from Millenium Nucleus NCN2023${\_}$002 (TITANs), ANID Millennium Science Initiative (AIM23-0001), and the China-Chile Joint Research Fund (CCJRF2310). PM acknowledges financial support from the Spanish MCIU through project PID2022-140871NB-C21 by “ERDF A way of making Europe”, and the Severo Ochoa grant CEX2021- 515001131-S funded by MCIN/AEI/10.13039/501100011033. SP is supported by the international Gemini Observatory, a program of NSF NOIRLab, which is managed by the Association of Universities for Research in Astronomy (AURA) under a cooperative agreement with the U.S. National Science Foundation, on behalf of the Gemini partnership of Argentina, Brazil, Canada, Chile, the Republic of Korea, and the United States of America.

\end{document}